\documentclass[aps,twocolumn,superscriptaddress,altaffilletter,lengthcheck,
tightenlines,showpacs,showkeys]{revtex4}
\newcommand{\al}{\alpha}
\newcommand{\bb}{\beta}
\newcommand{\D}{\Delta}
\newcommand{\ben}{\begin{eqnarray}}
\newcommand{\een}{\end{eqnarray}}
\newcommand{\be}{\begin{equation}}
\newcommand{\ee}{\end{equation}}
\newcommand{\ba}{\begin{eqnarray}}
\newcommand{\ea}{\end{eqnarray}}
\newcommand{\n}{\label}

\newcommand{\ga}{\gamma}
\newcommand{\ro}{\rho}
\newcommand{\om}{\omega}
\newcommand{\Om}{\Omega}

\newcommand{\bn}{\begin{equation}\label}
\usepackage{amssymb,amsmath}
\usepackage[dvipdf]{epsfig}
\usepackage{color}
\usepackage{float}

\usepackage{hyperref}

\begin{document}

\title{Self-interacting holographic dark energy}

\author{Luis P. Chimento}
\email{chimento@df.uba.ar}
\affiliation{Departamento de F\'isica, Facultad de ciencias Exactas y Naturales,
Universidad de Buenos Aires, 1428 Buenos Aires, Argentina}
\author{M\'onica Forte}
\email{monicaforte@fibertel.com.ar}
\affiliation{Departamento de F\'isica, Facultad de ciencias Exactas y Naturales,
Universidad de Buenos Aires, 1428 Buenos Aires, Argentina}
\author{Mart\'{i}n G. Richarte}\email{martin@df.uba.ar}
\affiliation{Departamento de F\'isica, Facultad de ciencias Exactas y Naturales,
Universidad de Buenos Aires, 1428 Buenos Aires, Argentina}

\begin{abstract}
 
We investigate a spatially flat Friedmann-Robertson-Walker (FRW) universe where dark matter exchanges energy  with a self-interacting holographic dark energy (SIHDE). Using the $\chi^2$--statistical method on the Hubble function, we obtain a critical redshift that seems to be consistent with both BAO and CMB data. We calculate the theoretical distance modulus for confronting with the observational data of SNe Ia for small redshift $z\leq 0.1$ and large redshift $0.1 \leq z\leq 1.5$.  The model  gets accelerate  faster than the $\Lambda$CDM one and  it can be a good candidate to alleviate the coincidence problem.  We also examine the age crisis at high redshift  associated with the old quasar APM 08279+5255.

\end{abstract}
\vskip .5cm


\date{\today}
\bibliographystyle{plain}

\maketitle

\section{Introduction}

 As it is well known our universe is currently undergoing an accelerated expansion phase driven by a mysterious fuel  called dark energy which exerts a negative pressure tending to drive clusters of galaxies apart. The latter fact has been corroborated by many different probes, for example the observation of  type Ia supernovae \cite{c1a,c1b,c2a,c2b},  the data of the large scale structure from SDSS \cite{c3a,c3b, c3c}, and  measurements of the cosmic microwave background anisotropy \cite{c4a,c4b,c4c}. The simplest candidate for the dark energy component is a positive cosmological constant $\Lambda$ \cite{lcdm1,lcdm2,lcdm3}. Although the prediction of the cold dark matter plus cosmological constant ($\Lambda$CDM) model is mostly consistent with  observational data, the  cosmological constant proposal suffers  from at least two puzzles  \cite{puzzleLambda1,puzzleLambda2,puzzleLambda3,puzzleLambda4,puzzleLambda5,puzzleLambda6,puzzleLambda7}. The first issue is  known as the fine-tuning problem, that is,  the theoretical prediction of the cosmological constant  that is obtained as the expectation value of quantum fields differs from its cosmic observed value by  120 orders of magnitude. The measured cosmological constant in our universe is tiny but not zero, and if it were much larger, galaxies could not have formed \cite{puzzleLambda4}. The second point of debate concerns the cosmic coincidence problem: why we observe that the fractional densities of dark matter and cosmological constant are about the same order of magnitude today.

 The conflict between theoretical physics and the observational data can be alleviated by working within the framework of dynamical dark energy \cite{dynaDEa,dynaDEb,dynaDEc}. This aforesaid idea has led to a wide variety of dark energy models such as quintessence \cite{quinta1, quinta2, quinta3, quinta4, quinta5, quinta6, quinta7, quinta8}, k-essence \cite{ke4,ke5,ke6,ke7}, quintom \cite{quinto1, quinto2, quinto3, quinto4, quinto6, quinto7, quinto8, quinto10, quinto11, quintomodi},  and  holography dark energy (HDE) \cite{hde0,hde1,hde2}. In particular, the latter model  was discussed  extensively during the last five years \cite{hde3, hde4,hde5, hde6,hde7,hde8, hde9, hde10, hde13, hde14, hde15a, hde15a, hde16a,hde16b,hde16c}.


The HDE model has its physical origin in the holographic principle as well as some features related with string and quantum gravity theories \cite{holop1, holop2, holop3, holop4, holop6}. The underlying postulate can be stated as follows \cite{holop6}: \textit{the number of degrees of freedom in a bounded system  should be finite and is related to the area of its boundary}. This principle also suggests that the ultraviolet (UV) cutoff scale of a system is
connected to its infrared (IR) cutoff scale. In the case of  a system  with size $L$ (IR length)  and ultraviolet   cutoff $\Lambda$ without decaying into a black hole, it is required that the total energy in the region of size $L$ should not exceed the mass of the black hole with the same size, thus,  $L^{3}\rho_{_{\Lambda}}\leq L M^{2}_{~P}$ being $M_{P}$ the reduced Planck mass  whereas  the  UV cutoff scale is defined as $\Lambda=\rho_{_{\Lambda}}^{1/4} $\cite{holop4}. The largest $L$ allowed is the one which saturates the above inequality and leads to  an holographic dark energy given by $\rho_{\Lambda}=3c^{2}M^{2}_{~P}L^{-2}$, where $c$ is a numerical factor. Hence,   this principle connects  the dark energy  based on the quantum zero-point energy density caused by a short distance cutoff $\Lambda$ with   an IR cutoff \cite{holop6} that  is usually taken  as the large scale of the universe, for instance, Hubble horizon \cite{hde0,hde1}, particle horizon \cite{hde1}, event horizon \cite{hde1} or generalized IR cutoff \cite{ricci1, ricci2, riccidiego, odiholo, ricciobse1, ricciobse2, ricciobse3, ricciobse4, diegoholoa,diegoholob, hmi1, hmi2}. 


 A natural arena for investigating   the coincidence problem is consider a phenomenological approach where dark matter interacts with  dark  energy \cite{acople1, acople2, acople3, quintaacople1, quintaacople2, quintaacople3, quintaacople4a,quintaacople4b, quintaacople5, quintaacople6, quintaacopleobse3}. From the observational point of view, an interacting dark sector is completely  compatible with the  current observations of standard candles and WAMP data \cite{quintaacopleobse1,quintaacopleobse2}.  In the present paper, we show how it is possible to get a physically viable model based on a new holographic dark energy density that  interacts with dark matter. More precisely, it turns to be  dark matter $\ro_c$ feels the presence of   dark energy $ \ro_x$  through the  gravitational expansion of the  universe plus  an exchange of energy between themselves.  Based   on the holographic principle,  we propose a dark energy  model where  the quantum zero point  energy density $\rho_{\Lambda}$ is equal to the dark energy density $\ro_x$ being $L$ an IR cutoff that will be related with a cosmological length. As a result of this,  we take $\ro_x =\rho_{\Lambda}=3c^{2}M^{2}_{~P}L^{-2}=3c^{2}M^{2}_{~P}f(\ro,p)$ where $f(\ro,p)$ is an arbitrary positive function. This gives rise to self-interacting holographic dark energy models (SIHDE) where $\ro_x\propto f(\ro,p)$, indicating that there is a coupling to the dark matter component.
The new  holographic dark energy model assumes a generalized IR cutoff  $L$  that depends on the total dark sector density $\ro=\ro_c+\ro_x$ and  the pressure of the mixture  $p=p_c+p_x$.

Several works have been devoted to obtain cosmological constraints in the case of Ricci scalar cutoff \cite{ricciobse1, ricciobse2,ricciobse3} or generalized versions of this one \cite{ricciobse4,hmi1,hmi2}. For example, the joint analysis of the 307 union sample of SNIa, together  with CMB shift parameter given by WMAP5, and the BAO measurement from SDSS, suggest that the holographic Ricci dark energy exhibits a quintom-like phase, so it leads to a new model consistent with the current observation because the equation of state for the Ricci dark energy can cross the phantom line \cite{ricciobse1}.
Using the general framework presented in \cite{jefe1}, which  suitably describes and unifies the dark sector with an exchange of energy, we will investigate a cosmological scenario where dark interacts with SIHDE. After that, we will confront our results with the current observational data and compare with the $\Lambda$CDM model. In the last section, we summarize
our main results and conclude.

\section{Evolution of the dark components }

We consider a flat FRW  universe filled with two components, dark matter and SIHDE with energy densities $\ro_c$ and $\ro_x$, respectively.  We also assume that the equations of states are  $\omega_c=p_c /\ro_c$ and $\omega_x=p_x/ \ro_x$,  whereas  the   Einstein equations read
\be
\n{01}
3H^2= \ro_c + \ro_x,
\ee
\be
\n{02}
\ro_c'+\ro_x'+(\omega_c+1)\ro_c + (\omega_x +1 )\ro_x=0.
\ee
Here $H = \dot a/a$ stands for the Hubble expansion rate, $a$ is the scale factor, and $ '$ means derivative with respect to the variable $\eta = \ln(a/a_0)^{3}$ being  $a_0$ the scale factor today. From Eqs. (\ref{01}) and (\ref{02}) the total pressure becomes, $p=-\ro'-\ro$, hence the SIHDE, $\ro_x=f(\ro,p)$ turns $\ro_x=f(\ro,\ro')$. As already mentioned in the introduction, $\ro_x$ is related with the UV cutoff, while $L=f^{-1/2}$ is related to the IR cutoff. We now consider the simplest case of a linear SIHDE,
\be
\n{03}
\ro_x=\frac{1}{\al -\bb}\left(\ro'+\al\ro\right),
\ee
where $\al$ and $\bb$ are both free constants. Rewriting  Eqs. (\ref{01}) and (\ref{03}) as
\be
\n{04}
\ro=\ro_c+\ro_x,
\ee
\be
\n{06}
\ro'=-\al\ro_c -\bb\ro_x,
\ee
and comparing   Eqs. (\ref{02})  with  Eq. (\ref{06}), we obtain a compatibility relation    
\be
\n{07}
\om_x=(\al-\om_c-1)r+\beta-1,
\ee
between the equation of state of both components and its ratio $r=\ro_c/\ro_x$. This relation allows us to use the Eq. (\ref{06}) with constant coefficients $\al$ and $\beta$ instead of the Eq. (\ref{02}) with  non-constant coefficients. After solving the linear system of equations (\ref{04}) and (\ref{06}), we obtain the energy density of each dark component as  functions of $\ro$ and $\ro'$
\be
\n{10}
\ro_c= - \frac{\bb \ro +\ro '}{\D}, \quad \ro_x= \frac{\al \ro +\ro '}{\D}, 
\ee
where $\D = \al -\bb$ is the determinant of the linear system of equations. At this point, we introduce  the interaction term, $3H Q_l$, between the dark components by splitting the Eq. (\ref{06}) in the following way
\be
\n{08}
\ro_c' + \al \ro_c = - Q_l,   \qquad \ro_x' + \bb \ro_x = Q_l.
\ee
After differentiating the first Eq. (\ref{10}) and combining with Eq. (\ref{08}), we find a second order differential equation for the total energy density:
\be
\n{14}
\ro''+(\al + \bb)\ro' + \al\bb\ro =  Q_l\D.
\ee
Once the interaction term $Q_l$ is selected and replaced in (\ref{14}), the total energy density $\ro$ of the dark sector is determined by solving the source equation (\ref{14}). Having obtained $\rho$, we are in position to get  $\ro_c$ and $\ro_x$  from Eq. (\ref{10}),   calculate the scale factor by integrating the Friedmann equation (\ref{01}), and find  the equation of state of the mixture from  the relation $p=-\ro'-\ro$. In the case of pressureless dark matter ($\om_c=0$),  the equation of state of dark energy  (\ref{07}) is given by 
\be
\n{gac1}
\om_x=(\al -1)r+(\beta-1),
\ee
so it becomes linear in $r$.
\section{Interacting holographic model}

In the present section, we are going to examine a proposal where the interaction term $Q_l$ is a general linear combination of $\ro_c$, $\ro_x$, $\ro$, and  $\ro'$ \cite{jefe1}
$$
Q_{l}= c_1 \frac{(\om_s +1 - \al)(\om_s +1-\bb)}{\Delta}\,\ro + c_2 (\om_s+1-\al)\ro_c-
$$
\begin{eqnarray} 
c_3 (\om_s +1 -\bb)\ro_x -c_4 \frac{(\om_s +1 - \al)(\om_s +1 -\bb)}{(\om_s+1) \Delta}\,\ro'\label{ql}.
\end{eqnarray}
Here $\om_s$ is a free constant parameter and  the coefficients $c_{i}$ fulfill the following condition
$c_{1}+c_{2}+c_{3}+c_{4}=1$ in order to assure the existence of  stable power law solutions $a=t^{2/3(\om_s +1)}$ \cite{jefe1}.
The case with $c_2=c_3=c_4=0$ was examined in \cite{hde13}, \cite{quintaacople2}, \cite{quintaacopleobse1}. The case $c_1=c_2=c_4=0$ was analyzed in  \cite{hdeIricci1, hdeIricci1ext, dedmdirection, dmdeintetermo}, \cite{interaDensDE1, interaDensDE2, interaDensDE3}. The linear interaction $Q_{l} \propto \ro'$, $c_1=c_2=c_3=0$, was introduced in  \cite{jefe1} and now it is considered here for its study.

Using Eqs. (\ref{10})  we can rewrite the interaction (\ref{ql}) as a linear combination of $\ro$ and $\ro'$ only, 
\be
Q_{l}=\frac{u\ro+(\om_{s}+1)^{-1}[u-(\om_s-\al+1)(\om_s-\beta+1)]\ro'}{\Delta},\label{qfinal}
\ee
where the parameter $u$ is  defined in terms of  $\om_s$, $\al$, and $\beta$ as follows:
$$u=c_1(\om_s -\al+1)(\om_s-\beta+1)$$
\be
\label{def.u}
\begin{split}
-c_{2}\beta(\om_s-\al+1)- c_{3}\al(\om_s-\beta+1).
\end{split}
\ee

Replacing the interaction term  (\ref{qfinal}) into the source equation (\ref{14}), we obtain a linear differential equation 
\begin{equation}
\ro'' + (\om_{s}+1)^{-1}[(\om_{s}+1)^{2} + \al\bb-u]\ro'+ (\al\bb-u)\ro=0,\label{se}
\end{equation}
whose characteristic polynomial roots are
\be
\n{r-+}
\ga^{-}=\om_{s}+1, \qquad \ga^{+}=\frac{\beta\al -u}{\om_s+1}. 
\ee
We restrict  our analysis to the case with positives roots in order to avoid phantom dark energy, then we choose $0<\om_s+1 < \ga^{+}$. Solving the Eq. (\ref{se}), we obtain  the total energy density in terms of the scale factor and consenquently the effective pressure:  
\begin{equation}
\ro=b_1a^{-3\ga^{+}}+b_2a^{-3(\om_s+1)},\label{densf}
\end{equation}
\be
\n{pb}
p= (\ga^{+}-1)b_1a^{-3\ga^{+}}+\om_{s} b_2a^{-3(\om_s +1)}.
\ee
From (\ref{10}) and (\ref{densf}), we get the dark matter and dark energy densities as a function of the scale factor
\be
\n{27}
\ro_c=\frac{(\ga^{+}-\beta)b_1a^{-3\ga^{+}}+(\om_s-\beta+1)b_2a^{-3(\om_s+1)}}{\Delta}, 
\ee
\be
\n{28}
\ro_x=\frac{(\al-\ga^{+})b_1a^{-3\ga^{+}}+(\al-\om_s-1)b_2a^{-3(\om_s+1)}}{\Delta}.
\ee
At very early times, dark matter and dark energy densities (\ref{27}) - (\ref{28}) behave as $a^{-3\ga^{+}}$ with a constant ratio $r_e\simeq  (\ga^{+}-\beta)/(\al-\ga^{+})$ while $p\simeq(\ga^{+}-1)b_1a^{-3\ga^{+}}$. However, at late times the effective fluid, dark matter, and dark energy have the same behavior with the scale factor, namely,  $\ro\simeq\ro_c\simeq\ro_x\simeq b_1 a^{-3(\om_s +1)}$, leading to  $r_l\simeq (\om_{s}-\beta+1)/(\al-1-\om_{s})$ and $p \simeq  a^{-3(\om_s +1)}$. The aforesaid facts indicate that the interaction term $Q_{l}$ is a good candidate to represent adequately an interacting dark sector because the ratio dark matter-dark energy $r$ has enough parameters to adjust the cosmological  observations and  it also alleviates the  so called coincidence problem.  It is important to emphasize that the mutual exchange of energy between the dark components makes that their usual behavior with scale factor change radically; we distingush in the dark densities two terms $a^{-3\ga^{+}}$ and $a^{-3(\om_s+1)}$.  In fact, we will consider the case with  $\ga^{+}=1$ in order to get pressureless dark matter at early times.


\section{Observational data analysis}

In this section, we will perform some qualitative cosmological constraints for the SIHDE model interacting with dark matter through the interaction term  $Q_{l}$ proposed in last section. In order to do that, we start by  constraining the parameter space with  the Hubble data $H(z)$ \cite{Stern:2009ep},\cite{Simon:2004tf}, and SNe Ia observations  \cite{Riess:2009pu}. The $H(z)$ test was probably first used to constrain cosmological parameters in \cite{Samushia:2006fx} and then in a large number of articles \cite{Wei:2006ut,Lazkoz:2007zk,Lin:2008wr,Cao:2011cg,Figueroa:2008py,Seikel:2012cs,Santos:2011cj,delCampo:2010zz,Aviles:2012ir,hmi1,hmi2,hologa,hologb,hologc}.
The statistical method requires the compilation of the observed value $H_{\rm obs}$ \cite{Stern:2009ep},\cite{Simon:2004tf}  and  the best value for the present time $z=0$ taken from \cite{Riess:2009pu}. Table \ref{Example} shows  $H_{\rm obs}$ at different redshift with its corresponding $1\sigma$ uncertainty and the reference where this value was reported. 
\begin{table}[htbp]
\begin{center}
\begin{tabular}{rccc}
\hline
$z$&$H(z)$&$1\sigma$&$\mbox{reference}$\\
$ $&${\rm km~s^{-1}\,Mpc^{-1}}$&$\mbox{uncertainty}$&$ $\\
\hline
0.000&73.8&$\pm  2.4$& \cite{Riess:2009pu}\\
0.090&69 &$\pm 12 $&\cite{Simon:2004tf}\\
0.170&83 &$\pm 8 $&\cite{Simon:2004tf}\\
0.179&75 &$\pm 4 $&\cite{Moresco:2012by}\\
0.199&75 &$\pm 5 $&\cite{Moresco:2012by}\\
0.270&77 &$\pm 14 $&\cite{Simon:2004tf}\\
0.352&83 &$\pm 14 $&\cite{Moresco:2012by}\\
0.400&95 &$\pm 17 $&\cite{Simon:2004tf}\\
0.480&97 &$\pm 62 $&\cite{Stern:2009ep}\\
0.593&104 &$\pm 13 $&\cite{Moresco:2012by}\\
0.680&92 &$\pm 8 $&\cite{Moresco:2012by}\\
0.781&105 &$\pm 12 $&\cite{Moresco:2012by}\\
0.875&125 &$\pm 17 $&\cite{Moresco:2012by}\\
0.880&90 &$\pm 40 $&\cite{Stern:2009ep}\\
1.037&154 &$\pm 20 $&\cite{Moresco:2012by}\\
1.300&168 &$\pm 17 $&\cite{Simon:2004tf}\\
1.430&177 &$\pm 18 $&\cite{Simon:2004tf}\\
1.530&140 &$\pm 14 $&\cite{Simon:2004tf}\\
1.750&202 &$\pm 40 $&\cite{Simon:2004tf}\\
\hline
\end{tabular}
\end{center}
\caption{\scriptsize{Hubble data $H_{\rm obs}(z_i)$ vs. redshift $z_i$ }}
\label{Example}
\end{table}
 
From Eqs. (\ref{01}), (\ref{27}), and (\ref{28}), we can write the Hubble function in terms of the effective equation of state $\om=\om_x\Om_x=\al\Om_c + \bb\Om_x=-2\dot H/3H^2$ as
follows
\be
\n{caso1}
H(z)=H_{0}\left[\frac{(\om_s-\om_0)(1+z)^3+\om_0(1+z)^{3(\om_{s}+1)}}{\om_s}\right]^{1/2},
\ee
where $\om_0=\al \Omega_{c0}+\bb \Omega_{x0}-1$, $\Omega_{x0}=\ro_{x0}/3H^{2}_{0}$, $\Omega_{c0}=\ro_{c0}/3H^{2}_{0}$ are their present values whereas the flatness condition today reads $\Omega_{c0}+\Omega_{x0}=1$. Taking into account  the transition  point $z_{acc}$, i.e. the  moment where the universe begins to accelerate  or where the deceleration parameter $q=-\ddot{a}/aH^2$ vanishes, the Eq. (\ref{caso1}) depends on  $(H_{0},z_{\rm acc},\om_s)$ parameters:
$$H(z)= \frac{H_0(1+ z_{\rm acc})^{3/2}} {\sqrt{-3\om_s-1+(1+ z_{\rm acc})^{-3\om_s}}}\times$$
\be
\n{Hdegazacc}
\left[\frac{(1+z)^{3(\om_{s}+1)}}{(1+z_{\rm acc})^{3(\om_{s}+1)}}-(1+3\om_s)\frac{(1+z)^3}{(1+z_{\rm acc})^3}\right]^{1/2}.
\ee
From Eq.(\ref{Hdegazacc}) we see that  the model has only three independent parameters $(H_{0},z_{\rm acc},\om_s)$ in order to be completely specified. The remaining parameters $\al$, $\beta$, $\Omega_{c0}$ or  $\Omega_{x0}$ are included in the transition point $z_{acc}$ through the equation of state $\om_0$. 
 We now  proceed in the following way: we  perform a statistical analysis on the $(H_{0},z_{acc},\om_s)$  parameters, confronting their best fit values with the recent available data and then, we will perform the same Hubble test using the expression (\ref{caso1}), to obtain constraints on the others parameters, $\al$, $\beta$, and $\Omega_{c0}$. The second approach will give us the most favored  SIHDE  for the Hubble's data.

The probability distribution for the $\theta$-parameters is $P(\theta)=\mathcal N e^{-\chi^2(\theta)/2}$ (see e.g. \cite{Press}) being $\mathcal N$  a normalization constant. The parameters of the model are estimated by minimizing the $\chi^{2}$ function of the Hubble data which is constructed as 
\be
\chi^2(\theta) =\sum_{i=1}^{N=19}\frac{[H(\theta;z_i) - H_{\rm obs}(z_i)]^2}{\sigma^2(z_i)},
\ee 
where $\theta$ stands for the cosmological parameters, $H_{\rm obs}(z_k)$ is the observational $H(z)$ data at the redshift $z_k$, $\sigma(z_k)$ is the corresponding $1\sigma$ uncertainty, and the summation is over the $19~$ observational  $H(z)$ data. The Hubble function is not integrated over and it is directly related with the properties of the dark energy, since its value comes from the cosmological observations. Using the absolute ages of passively evolving galaxies observed at different redshifts, one obtains the differential ages $dz/dt$ and the function $H(z)$ can be measured through the relation $H(z)=-(1+z)^{-1}dz/dt$.  The $\chi^2$ function reaches its  minimum value at the best fit value $\theta_c$ and the fit is good when $\chi^2_{\rm min}(\theta_c)/(N -n) \leq 1$, where $n$ is the number of parameters \cite{Press} and  $N$ counts the observational data points that in our case correspond to  $19$ points.

\begin{figure}[h!]
\begin{center}
\includegraphics[height=9cm,width=9cm]{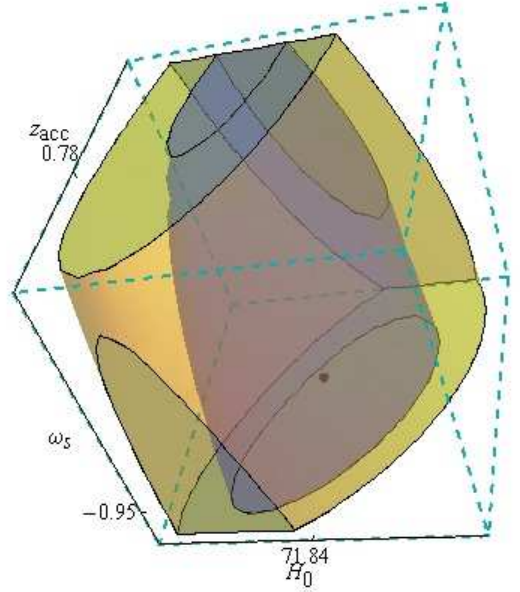}
\caption{\scriptsize{Three-dimensional C.L. assocaited with $1\sigma$, $2\sigma$ for $H_0 $, $\om_s$,   and  $z_{\rm acc}$ parameters. The point indicates the best fit observational values, namely,    $H_0= 71.839~{\rm km s^{-1}\,Mpc^{-1}}$,  $ z_{\rm acc}=0.7831$ and $\om_s= - 0.950$.}}
\label{Fig:. Caso II,dim3}
\end{center}
\end{figure}
\begin{figure}[h!]
\begin{center}
\includegraphics[height=9cm,width=9cm]{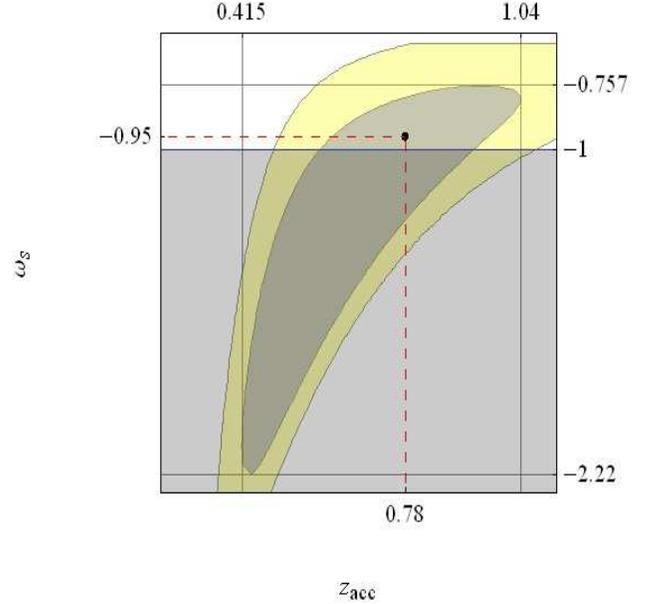}
\caption{\scriptsize{Two-dimensional C.L. assocaited with $1\sigma$, $2\sigma$ for $\om_s$    and  $z_{\rm acc}$ parameters, after the  marginalization over over the parameter $H_0 $ was done. The point indicates the best fit observational value obtained with the $H(z)$ function.}}
\label{Fig:. Caso II,dim2Marginalized}
\end{center}
\end{figure}
\begin{figure}[h!]
\begin{center}
\includegraphics[height=8cm,width=9cm]{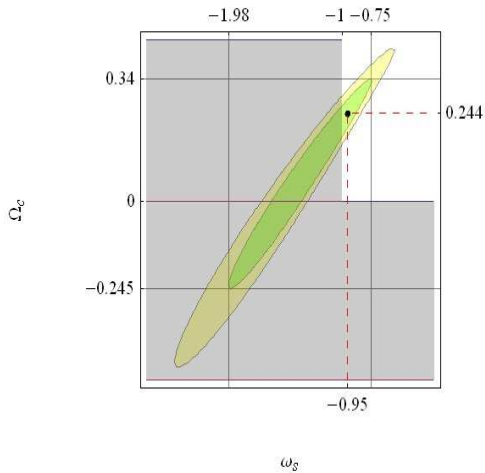}
\caption{\scriptsize{Constraints in the  $\omega_s $-$\Omega_{c0}$ plane. Elliptical two-dimensional  C.L. associated with $1\sigma$ and $2\sigma$ error bars. The dot indicates the best fit observational value obtained with the $H(z)$ function. The shaded zone  is  excluded because we try to avoid phantom dark energy in our model.}}
\label{Fig:. dim2}  
\end{center}
\end{figure}

In the first approach,  the  parameters of the model are $\theta=(H_0, z_{acc},\om_s)$ therefore the  $68.3\%$ ($1\sigma$) or $95.4\%$ ($2\sigma$)  confidence levels (C.L.) made with  the random data fulfill the inequalities $\chi^{2}(\theta)-\chi^{2}_{\rm min}(\theta_c)\leq 3.53$ or $\chi^{2}(\theta)-\chi^{2}_{\rm min}(\theta_c)\leq 8.02$, respectively.  Fig.\ref{Fig:. Caso II,dim3} shows the C.L. associated with  $1\sigma$ and $2\sigma$ error bars in the $H_0- z_{\rm acc}-\om_s$ space; we find  the best-fit values at $H_0= 71.839~{\rm km s^{-1}\,Mpc^{-1}}$,  $ z_{\rm acc}=0.7831$ and $\om_s= - 0.95$ corresponding to a $ \chi^2_{\rm min}=14.26$ along with  $\chi^2_{\rm d.o.f}=\chi^2_{\rm min}/(N-n)=0.790$ per degree of freedom. We remark that  our estimations of the actual Hubble parameter agree with  the median statistics made in  \cite{Chen:2011ab}, namely,  our value   meets within  the $1\sigma$ interval  obtained with the median statistics,  $H_{0}=68 \pm 5 ~{\rm km s^{-1}\,Mpc^{-1}}$,   or with the analysis performed in \cite{Calabrese:2012vf}
 about the impact of $H_{0}$ prior on the evidence for dark radiation. On the other hand,  we obtain C.L. in the  $z_{\rm acc}-\om_s$ plane  obtained after having marginalized the joint probability $P(H_0, z_{\rm acc}, \om_s)$ over $H_0$ (see Fig.\ref{Fig:. Caso II,dim2Marginalized}).  As usual, in the case of two parameters,  $68.3\%$, $95.4\%$ C.L. are made of random data sets that satisfy  the inequality  
 $\chi^{2}(\theta)-\chi^{2}_{\rm min}(\theta_c)\leq 2.3$,  $\chi^{2}(\theta)-\chi^{2}_{\rm min}(\theta_c)\leq 6.17$, respectively \cite{Press}. The shaded band corresponding to $ \om_s \leq -1$ is excluded in our model in order to avoid phantom dark energy. The constraint on the critical redshift is $z_{\rm acc}=0.78 ^{+ 0.26}_{~-0.37}$, such value  are in agreement with $z_{t}=0.69^{+0.20}_{-0.13}$ reported in \cite{Zt1}-\cite{Zt2}, and meets  within the $2 \sigma$ C.L  obtained with the supernovae (Union 2) data in \cite{Zt2}. The critical redshift   $z_{\rm acc}=0.78 ^{+ 0.26}_{~-0.37}$ is also consistent with  ${\rm Union~ 2+BAO+CMB}$  data \cite{Li:2010da}. For the other parameter the statistical analysis leads to $\om_{s}=-0.95^{+0.20}_{~-1.27}$. 

In order to get some  physically relevant bounds on  $\al$, $\beta$,  and $\Omega_{c0}$ parameters, we now use the expression (\ref{caso1}) and take as prior $H_0= 71.84~{\rm km s^{-1}\,Mpc^{-1}}$ which is  in agreement  with  the median statistical constraints found in \cite{Chen:2011ab}-\cite{Calabrese:2012vf}. Taking into account  (\ref{caso1}) for the $\chi^2$ statistical analysis, we obtain the best-fit values $\al=1.15$, $\beta=0.023$,  and $\Omega_{c0}=0.243$ with $ \chi^2_{\rm min}=14.2623$ along with  $\chi^2_{\rm d.o.f}=0.95<1$.   Fig.\ref{Fig:. dim2} shows  two-dimensional C.L.  in  the $\om_s$-$\Omega_{c0}$ plane whereas the other parameters are taken as priors, namely, we fix $\al=1.15$, $\beta=0.023$, and $H_0= 71.84~{\rm km s^{-1}\,Mpc^{-1}}$. Then, the best-fit values together with their error bars are  $\om_{s}=-0.95 ^{+ 0.20}_{~-1.03}$ and $\Omega_{c0}=0.244 ^{+ 0.096}_{~-0.489}$.  
\begin{figure}[h!]
\begin{center}
\includegraphics[height=7cm,width=7cm]{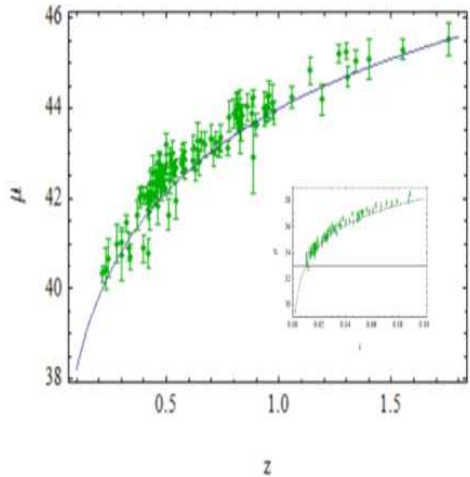}
\caption{\scriptsize{The plot of the theoretical distance modulus (solid line) versus the redshift. The observational data (point) was taken from Riess \cite{Riess:2009pu} and covers two different regions of redshift, $ z \leq 0.1 $ and $0.1 \leq z \leq 1.5$. We obtain that the best-fit values, obtained for the interacting model, are in agreement with the  supernovae data.}}
\label{Fig:. Caso II,modulo}
\end{center}
\end{figure}
We would like to use the best-fit values  $\om_s = -0.95$ and $\Omega_{c0}=0.244$ to calculate the   magnitude redshift relation for standard candles and contrast with  the supernova data. As it is well known the observations of SNe Ia have predicted  and   confirmed that our universe is currently passing through an accelerated phase of expansion. Since then, the observational data  coming from these standard candles have been taken  very seriously.  It is commonly believed that by measuring both their redshifts and apparent peak flux gives a direct measurement of their luminosity distances and thus SNe Ia provide the strongest constraint on the cosmological parameters. The theoretical distance modulus is defined as 
\be
\n{mut}
\mu(z) =5~\log_{10} {\cal D}_{L}+\mu_{0},
\ee
where $\mu_{0}=43.028$, and ${\cal D}_{L}$ is the Hubble-free luminosity distance, which for a spatially flat universe can be recast as 
\be
\n{mut'}
{\cal D}_{L}(z) =(1+z)H_{0}\int^{z}_{0}\frac{dz'}{H(z')}.
\ee
Replacing the best-fit values of $H_0$, $\om_s$, and $\Omega_{c0}$ in Eqs.(\ref{caso1})-(\ref{mut'}) we get the theoretical distance modulus  $\mu(z)$ for our model (see Fig. \ref{Fig:. Caso II,modulo}) whereas the observational data with their error bars, $\mu_{obs}(z_i)$, are taken from  \cite{Riess:2009pu}. As we can see from Fig.\ref{Fig:. Caso II,modulo}, our model exhibit  an  excellent agreement with the observational data, at least in the zones corresponding to small redshifts [$ z \leq 0.1 $] and large  redshifts [$0.1 \leq z \leq 1.5$]. 

\subsection{The age problem}

We now  turn our attention to the age problem, namely,  the universe cannot be younger than its constituents (see \cite{ap1}). For example, the matter-dominated FRW universe can be ruled out because its age is smaller than the ages inferred from old globular clusters. The age problem becomes even more serious when we consider the age of the universe at high redshift. Now, there are some old high redshift objects (OHROs) discovered, for instance, the 3.5 Gyr old galaxy LBDS 53W091 at redshift $z = 1.55$ \cite{ap2a,ap2b}, the 4.0 Gyr old galaxy LBDS 53W069 at redshift $z = 1.43$ \cite{ap3}, the 4.0 Gyr old radio galaxy 3C 65 at $z = 1.175$ \cite{ap4}, and the high redshift quasar B1422+231 at $z = 3.62$ whose best-fit age is 1.5 Gyr with a lower bound of 1.3 Gyr \cite{ap5}. Also the old quasar APM 08279+5255 at $z = 3.91$, whose age is estimated to be $2.0-3.0$ Gyr \cite{ap6a,ap6b}, is used extensively. To assure the robustness of our analysis, we use the most conservative lower age estimate 2.0 Gyr for the old quasar APM 08279+5255 at $z = 3.91$ \cite{ap6a, ap6b}, and the lower age estimate 1.3 Gyr for the high redshift quasar B1422+231 at $z = 3.62$ \cite{ap5}. Many authors have examined   the age problem within the framework of the dark energy models, see e.g. \cite{ap1}, \cite{ap7}-\cite{ap13a,ap13b}, and references therein. The age problem within the context of holographic dark energy model was  explored in \cite{ap12} and \cite{ap14a, ap14b, ap14c}. In this section, we would like to consider the age problem for the SIHDE model with linear interaction. 

The age of our universe at redshift $z$ can be obtained from the dimensionless age parameter (\cite{ap1}, \cite{ap8})
\be
\n{atdez}
T_{z}(z)=H_{0}t(z)=H_{0}\int^{\infty}_{z}\frac{dz'}{(z'+1)H[z']}.
\ee
At any redshift, the age of our universe should be larger or equal than the age of the old high redshift objects
\be
\n{destdez}
T_{z}(z) \geq T_{obj} =H_{0}t_{obj}, ~\mbox{or}~ S(z)=\frac{T_{z}(z)}{T_{obj}} \geq 1
\ee
where $t_{obj}$ is the age of the OHRO. It is worth noting that from Eq. (\ref{atdez}), $T_{z}(z)$ is independent of the Hubble constant $H_0$. On the other hand, from Eq. (\ref{destdez}), $T_{obj}$ is proportional to the Hubble constant $H_0$ that we consider as $H_0= 71.84~{\rm km s^{-1}\,Mpc^{-1}}$.

In Table \ref{age}, we show the ratio $S(z)=T_{z}(z)/T_{obj}$ at $z=3.91, 3.62, 1.55, 1.43, 1.175$  taking to account the best-fit values obtained in the last section. We obtain that $T_{z}(z)>T_{obj}(z)$  at $z= 3.62, 1.55, 1.43, 1.175$ but  $T_{z}(z)<T_{obj}(z)$ at $z=3.91$, so the old quasar APM 08279+5255 cannot be accommodated as the others old objects. Perhaps, the age crisis at high redshift in the case of dark energy holographic models \cite{ap12}, \cite{ap14} could be alleviated by taking into account another type of interaction. This fact will be explored in a future research.

\begin{table}[htbp]
\begin{center}
\begin{tabular}{|ccccc|c|}
\hline
 $S(3.91)$ &$S(3.62)$&  $S(1.55)$ &$S(1.43)$&$S(1.175)$
\\ \hline
0.854555&1.19781&1.20467&1.12826&1.31645\\
\hline
\end{tabular}
\end{center}
\caption{\scriptsize{It shows the ratio $S(z)=T_{z}(z)/T_{obj}$ at $z=3.91, 3.62, 1.55, 1.43, 1.175$  for the best-fit values obtained with the Hubble data. }}
\label{age}
\end{table}
\subsection{Kinematic analysis}

\begin{figure}[h!]
\begin{center}
\includegraphics[height=9.3cm,width=8cm]{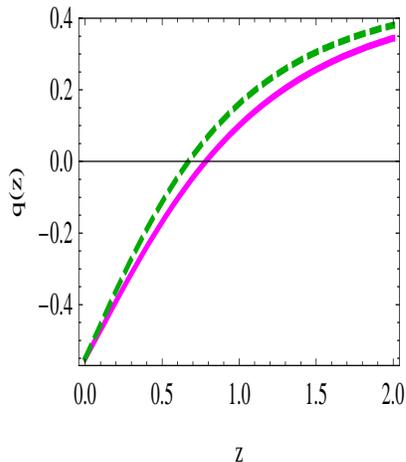}
\caption{\scriptsize{Plot of deceleration parameter $ q(z)$ taking into account the best-fit values $ \Omega_x =0.75$, $\al=1.15$, $\bb=0.024$, and $\om_s=-0.95$ (magenta, solid line). It also shows the deceleration parameter for the $\Lambda CDM$ model (green, dashed line).}}
\label{qComparados}
\end{center}
\end{figure}

Fig.\ref{qComparados} shows the behavior of  the deceleration parameter  with redshifts. Using the values $\Omega_{c0}=0.244$ and $\Omega_{x0}=0.75$, we obtain that  the deceleration parameter vanishes at $z_{\rm acc} =0.78$, so the universe enters the accelerated phase earlier  than the $\Lambda$CDM model.  Regarding the effective equation of state, it stays in the range $-1<\omega(z)<0$ for  $z\geq 0$, more precisely, $\omega(z)$ starts as non relativistic  cold matter, decreases rapidly around $z=2$ and then ends with the asymptotic value $\omega_{s}=-0.95$. The dark energy equation of state $\omega_{x}$ stays in the range $-1<\om_x (z)<0$ also.
The ratio dark matter-dark enery, 
\be
\n{rdez}
r=\frac{\om_{0}(\beta-\om_s-1)+(1-\beta)(\om_{0}-\om_{s})(1+z)^{3\om_{s}}}{\om_{0}(\om_s+1-\al)+(\al-1)(\om_{0}-\om_{s})(1+z)^{3\om_{s}}},
\ee
evaluated at the best fit values $\al=1.15$, $\bb=0.024$, $\om_s=-0.95$ and $\om_{0}=0.78$ indicates that interaction $Q_{l}$ helped to alleviate the coincidence problem. 

\section{Summary and Conclusions}

In this paper we have considered  a flat FRW universe composed of an interacting dark matter and  SIHDE. We have shown that the compatibility between SIHDE and the conservation equation gives a constraint between  the equations of state of the dark components. We have selected  linear SIHDE and a linear interaction in the dark sector and find that this model describes properly the evolution of both dark components. We have also shown that  a general linear  interaction, $Q_{l}$, is  a good candidate for alleviating  the cosmic coincidence problem.

Taking into account the Hubble data (see Table \ref{Example}) and using the $\chi^{2}$ statistical method, we  have obtained the best-fit values at $H_0= 71.839~{\rm km s^{-1}\,Mpc^{-1}}$,  $ z_{\rm acc}=0.7831$ and $\om_s= - 0.95$  along with  $\chi^2_{\rm  d.o.f}=0.790<1$ per degree of freedom (see Fig. \ref{Fig:. Caso II,dim3}). The value of  $H_{0}$  is in agreement with the one  reported in the literature \cite{Riess:2009pu} or with  the median statistical constraints found in \cite{Chen:2011ab}, \cite{Calabrese:2012vf}.  Having marginalized the joint probability $P(H_0, z_{\rm acc}, \om_s)$ over $H_0$ (see Fig.\ref{Fig:. Caso II,dim2Marginalized}) we  build two dimensional C.L. and  obtained  the best-fit values with their $1\sigma$ error bars, namely,   $z_{\rm acc}=0.78 ^{+ 0.26}_{~-0.37}$ and $\om_{s}=-0.95^{+0.20}_{~-1.27}$. The critical redshift is in agreement with $z_{t}=0.69^{+0.20}_{-0.13}$ reported in \cite{Zt1}-\cite{Zt2}, and meets  within the $2 \sigma$ C.L  obtained with the supernovae (Union 2) data in \cite{Zt2}. It is also consistent with  ${\rm Union~ 2+BAO+CMB}$  data \cite{Li:2010da}. Using as priors $\al=1.15$, $\beta=0.023$, and $H_0= 71.84~{\rm km s^{-1}\,Mpc^{-1}}$, we build two-dimensional C.L.  in  the $\om_s$-$\Omega_{c0}$ plane and estimated the best-fit values  $\om_{s}=-0.95 ^{+ 0.20}_{~-1.03}$ and $\Omega_{c0}=0.244 ^{+ 0.096}_{~-0.489}$  (see Fig.\ref{Fig:. dim2} ).

Replacing the best-fit values of $H_0$, $\om_s$, and $\Omega_{c0}$, we obtained the theoretical distance modulus  $\mu(z)$ for our model (see Fig. \ref{Fig:. Caso II,modulo}) and confronted with supernovae data $\mu_{obs}(z_i)$ taken from  \cite{Riess:2009pu}. Fig. \ref{Fig:. Caso II,modulo} shows that our model exhibit  an  excellent agreement with the observational data. Besides, we have found that the age crisis at high redshift cannot be alleviated  because the old quasar APM 08279+5255 at $z = 3.91$ \cite{ap6a,ap6b} seems to be older than the universe;  so it will be needed to consider other kind of interaction (cf. Table \ref{age}) or perhaps to propose  a non linear SIHDE for exploring this issue.  Finally, we have found that our model enters the accelerated phase faster than the $\Lambda$CDM model (see Fig.\ref{qComparados}). Concerning  the effective equation of state and the dark energy equation of state, we found that both do not cross the phantom divide line \cite{quinto7}. In a future research, we are going to explore the linear SIHDE proposal where the dark sector is also coupled to  a radiation or baryonic term;  we will examine the changes introduced in the behavior of dark energy at early times \cite{tfmi}.

\acknowledgments
We are grateful with the referee for useful comments that helped improve the article.
L.P.C thanks  the University of Buenos Aires under Project No. 20020100100147 and the Consejo Nacional de Investigaciones Cient\'{\i}ficas y T\' ecnicas (CONICET) under Project PIP 114-200801-00328. M.G.R is partially supported by CONICET.


\end{document}